# Diamagnetic d-orbitals Drive Magnetic Structure Selection in the Double Perovskite Ba$_2$MnTeO$_6$

Otto H. J. Mustonen,[1] Charlotte E. Pughe,[1] Helen C. Walker,[2] Heather M. Mutch,[1] Gavin B. G. Stenning,[2] Fiona C. Coomer,[3] Edmund J. Cussen[1]*

[1]Department of Material Science and Engineering, University of Sheffield, Mappin Street, Sheffield S1 3JD, United Kingdom

[2]ISIS Pulsed Neutron and Muon Source, STFC Rutherford Appleton Laboratory, Harwell Campus, Didcot OX11 0QX, United Kingdom

[3]Johnson Matthey Battery Materials, Blount's Court, Sonning Common, Reading RG4 9NH United Kingdom

* corresponding author:
Edmund J. Cussen
e.j.cussen@sheffield.ac.uk

B-site ordered A$_2$B'B''O$_6$ double perovskites have a variety of applications as magnetic materials. Here we show that diamagnetic d$^{10}$ and d$^0$ B'' cations have a significant effect on the magnetic interactions in these materials. We present a neutron scattering and theoretical study of the Mn$^{2+}$ double perovskite Ba$_2$MnTeO$_6$ with a 4d$^{10}$ Te$^{6+}$ cation on the B''-site. It is found to be a Type I antiferromagnet with a dominant nearest-neighbor J$_1$ interaction. In contrast, the 5d$^0$ W$^{6+}$ analogue Ba$_2$MnWO$_6$ is a Type II antiferromagnet with a significant next-nearest-neighbor J$_2$ interaction. This is due to a d$^{10}$/d$^0$ effect, where the different orbital hybridization with oxygen 2p results in different superexchange pathways. We show that d$^{10}$ B'' cations promote nearest neighbor and d$^0$ cations promote next-nearest-neighbor interactions. The d$^{10}$/d$^0$ effect could be used to tune magnetic interactions in double perovskites.

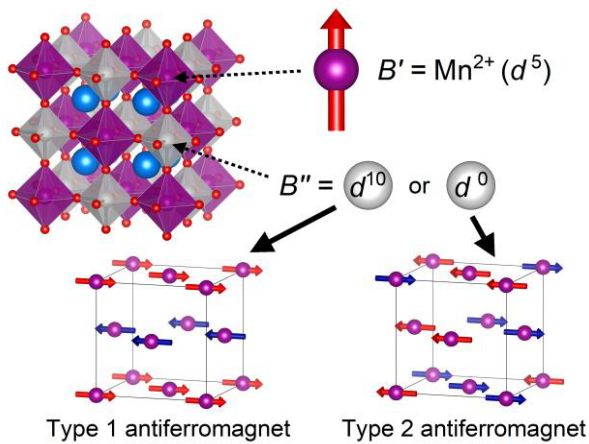

INTRODUCTION

The B-site ordered $A_2B'B''O_6$ double perovskite family of oxides has been widely investigated for their varied functional properties. In terms of magnetic materials, double perovskites have potential applications as ferro- and ferrimagnets with high Curie temperatures, as half-metallic ferromagnets in spintronics and as multiferroic materials.[1] The magnetic ground state and transition temperature of a material are determined by the interactions between its magnetic cations. In oxides, the main interactions are usually superexchange interactions mediated by oxygen 2p orbitals. The semi-empirical Goodenough-Kanamori rules[2–4] predict superexchange interactions in materials, where two transition metal cations are separated by an oxygen anion in either a 90° or a 180° bond angle.

The situation is more complicated in compounds where the magnetic cations are further apart. This occurs in double perovskites with one magnetic and one diamagnetic B-site cation alternating over the sites of the structure giving rock-salt order. Diamagnetic $d^{10}$ and $d^0$ cations can act as linkers between magnetic cations in extended superexchange pathways, and can lead to very different magnetic interactions. This is due to an orbital hybridization mechanism, where the empty $d^0$ orbitals can readily hybridize with O 2p orbitals whereas the full $d^{10}$ states tend to be significantly below the Fermi level.[5] This $d^{10}/d^0$ effect has been firmly established in B-site ordered $Cu^{2+}$ double perovskites $Sr_2Cu(Te,W)O_6$ and $Ba_2Cu(Te,W)O_6$, where $Te^{6+}$ is a $4d^{10}$ cation and $W^{6+}$ is a $5d^0$ cation on the B''-site linking the $CuO_6$ octahedra.[6–13] Despite being isostructural, the interactions in the $Te^{6+}$ and the $W^{6+}$ compounds are completely different: both $Sr_2CuTeO_6$ and $Ba_2CuTeO_6$ have dominant nearest-neighbor $J_1$ interactions whereas $Sr_2CuWO_6$ and $Ba_2CuWO_6$ have dominant next-nearest-neighbor $J_2$ interactions.[8,10] Furthermore, a novel spin-liquid-like state has been reported for the solid solution $Sr_2CuTe_{1-x}W_xO_6$.[13–16]

These $Cu^{2+}$ compounds are not representative of double perovskites in general for a number of reasons. The $Cu^{2+}$ double perovskites are tetragonally distorted by the Jahn-Teller activity of this cation resulting in more complicated interactions. Octahedral tilts are closely linked to the overall strength of magnetic interactions in these materials.[17] Most importantly, due to orbital ordering of the $3d^9$ $Cu^{2+}$ cations, the magnetism in these materials is highly two-dimensional unlike in other double perovskites.[8,10,17] The interactions only involve the unoccupied $3d_{x^2-y^2}$ orbitals of the copper cations. Finally, quantum effects such as quantum fluctuations are significant because of S = 1/2 nature of the $Cu^{2+}$ cation. Therefore, it remains an open question whether the $d^{10}/d^0$ effect is specific to these $Cu^{2+}$ compounds or whether it is general for all double perovskites. Differences between materials with $d^{10}$ or $d^0$ linking cations have also been reported for some osmate double perovskites, but there the situation is complicated by the strong spin-orbit coupling of 5d cations.[18–21]

The $Mn^{2+}$ double perovskites $Ba_2MnTeO_6$ and $Ba_2MnWO_6$ are ideal materials for testing the $d^{10}/d^0$ effect. Both crystallize in the cubic $Fm\bar{3}m$ space group, where the magnetic $Mn^{2+}$ cations form an undistorted fcc lattice. The lattice parameters and bond distances are almost identical due to the similar ionic radii of the $4d^{10}$ $Te^{6+}$ and $5d^0$ $W^{6+}$ cations.[22] Magnetic interactions in these perovskites are three-dimensional unlike in the copper compounds, and there are no quantum effects due to the large, classical-like spin-5/2 on $3d^5$ $Mn^{2+}$ cations. Moreover, there is no spin-orbit coupling nor an orbital moment. As a result, magnetism in both compounds can be described using a simple fcc Heisenberg model with only two interactions as shown in Figure 1: nearest neighbor $J_1$ and next-nearest-neighbor $J_2$ interaction. We have recently reported the full magnetic characterization of $Ba_2MnWO_6$.[23] The material is cubic with full ordering of the $Mn^{2+}$ and $W^{6+}$ cations on the B-sites. The space group is $Fm\bar{3}m$ with a = 8.18730(2) at 2 K. The Néel temperature of $Ba_2MnWO_6$ is 8 K, and it has a Type II antiferromagnetic structure. The $J_1$ interaction is slightly stronger than the $J_2$ with $J_1$ = -0.080 meV and $J_2$ = -0.076 meV, respectively.

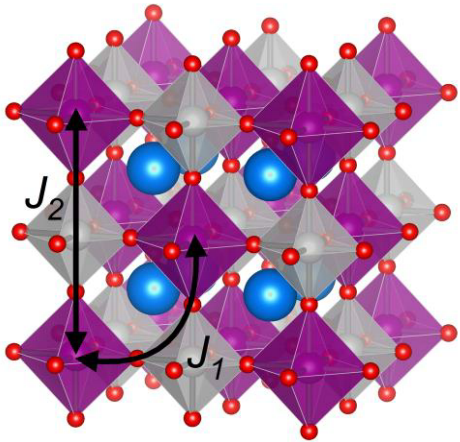

Figure 1. The B-site ordered double perovskite structure of $Ba_2MnTeO_6$ and $Ba_2MnWO_6$. The A, B′, B′′ and O sites are represented by blue, violet, gray and red spheres, respectively. The main magnetic interactions are shown: nearest-neighbor $J_1$ (corner to face center) and next-nearest neighbour $J_2$ (corner to corner).

In this article, we present the magnetic properties of the $Mn^{2+}$ double perovskite $Ba_2MnTeO_6$ for the first time. $Ba_2MnTeO_6$ has been synthesized before and a structure was proposed, but its physical properties have not been investigated.[24] It is a $4d^{10}$ $Te^{6+}$ analogue of $Ba_2MnWO_6$, and allows us to study the effect of $d^{10}$ and $d^0$ diamagnetic cations on the superexchange interactions in double perovskites. $Ba_2MnTeO_6$ is isostructural with $Ba_2MnWO_6$, and crystallizes in the space group $Fm\bar{3}m$ with a = 8.2066(3) at 2 K. Despite minimal differences in crystal structure including bond distances and angles (< 0.5%), the magnetic properties of $Ba_2MnTeO_6$ and $Ba_2MnWO_6$ are very different. Neutron diffraction experiments reveal a Type I antiferromagnetic structure in $Ba_2MnTeO_6$ below $T_N$ = 20 K, whereas $Ba_2MnWO_6$ has a Type II magnetic structure. The main superexchange interaction in $Ba_2MnTeO_6$ is found to be the $J_1$ = -0.34 meV, while the $J_2$ interaction is very weak with $J_2$ = 0.03 meV.

The significant differences in exchange interactions and magnetic structures between these two isostructural compounds can be understood in terms of differences in orbital hybridization between the $d^{10}$ and $d^0$ cations. We conclude that $d^{10}$ cations on the B″ site in double perovskites favor antiferromagnetic $J_1$ interactions whereas $d^0$ cations enhance antiferromagnetic $J_2$ interactions. This is supported by a survey of known ordered double perovskites with either $d^{10}$ $Te^{6+}$ or $d^0$ $W^{6+}/Mo^{6+}$ cations on the B″-site. This $d^{10}/d^0$ effect could be used to tune magnetic interactions and ground states in double perovskites in general as has been demonstrated for $Sr_2CuTe_{1-x}W_xO_6$.[13–16] Finally, as in $Ba_2MnWO_6$, we also observe a short-range correlated magnetic state in $Ba_2MnTeO_6$ that survives up to T ~ $5T_N$.

EXPERIMENTAL SECTION

Polycrystalline powder samples of $Ba_2MnTeO_6$ were prepared using a conventional solid-state chemistry method. Stoichiometric quantities of $BaCO_3$ (99.997%), $MnO_2$ (99.999%) and $TeO_2$ (99.9995%) were mixed in an agate mortar, pelletized and calcined at 900 °C in air. The synthesis was carried out in air at 1100 °C for 96 h with intermittent grindings. Phase purity was investigated using X-ray diffraction (Rigaku Miniflex, Cu Kα).

Magnetic properties were measured on a Quantum Design MPMS3 magnetometer. 111.94 mg of sample powder was enclosed in a gelatin capsule, immobilized with PTFE tape and then placed in a straw sample holder. The measurements were performed in SQUID-VSM mode in the temperature range 2-300 K in an applied field of $\mu_0H$ = 0.1 T. Both zero-field and field-cooled measurements were taken. Specific heat was measured with a Quantum Design PPMS instrument using a thermal relaxation method. A 6.84 mg pellet piece was used for the measurements, which were carried out between 2-60 K.

The nuclear and magnetic structure of $Ba_2MnTeO_6$ was investigated using neutron diffraction. The measurements were performed on the GEM time-of-flight diffractometer at the ISIS Neutron and Muon Source.[25] 7.9 g of sample powder was enclosed in a vanadium can. The collected data is available online.[26] Rietveld refinement was carried out using FULLPROF.[27] The crystal structures were visualized using VESTA.[28]

Magnetic excitations in $Ba_2MnTeO_6$ were measured using time-of-flight inelastic neutron scattering (INS) at the MERLIN instrument[29] at the ISIS Neutron and Muon Source. The same 7.9 g powder sample was used for both the INS and diffraction experiments. The sample powder was enclosed in an aluminium foil sachet to form an annulus inside a cylindrical aluminium can, which was inserted into a closed-cycle refrigerator to measure between 7 and 200 K. Measurements were performed using a Gd chopper and rep-rate multiplication[30–32] to allow S(Q,E) dynamical structure maps as a function of momentum and energy transfer to be recorded simultaneously with incident energies of 10, 20 and 53 meV. The data were reduced using the MantidPlot software package.[33] The raw data were corrected for detector efficiency and time independent background following standard procedures. The datasets are available online at ref. [34].

The electronic structures of $Ba_2MnTeO_6$ and $5d^0$ $W^{6+}$ analogue $Ba_2MnWO_6$ were calculated using density functional theory. The calculations were carried out using the full-potential linearized plane-wave (FPLAPW) code ELK[35]. The calculations were performed with generalized gradient approximation functionals by Perdew, Burke and Ernzerhof[36] (GGA-PBE) with a plane-wave cut-off of $|G + k| = 8/R_{MT}$ a.u.$^{-1}$, where $R_{MT}$ is the average muffin-tin radius. Electron correlation effects of the localized Mn 3d orbitals were included in a DFT+U approach with the Hubbard U and Hund coupling J as parameters.[37] The calculations were performed with U = 5 and 7 eV with J = 0.9 eV. We used the experimental crystal structures in the calculations with a Type I antiferromagnetic structure, which can be calculated without creating a supercell. While this is not the correct magnetic structure for $Ba_2MnWO_6$, it has little effect on the density of states, which is our main interest.

RESULTS

Magnetic properties and specific heat

Magnetic properties of $Ba_2MnTeO_6$ were investigated using DC magnetometry (Figure 2). Two transitions are observed in the zero-field cooled (ZFC) curve: an antiferromagnetic transition at ~20 K and a ferro- or ferrimagnetic transition at ~45 K. The field cooled (FC) curve diverges from the ZFC curve at the latter transition. The antiferromagnetic transition is inherent to $Ba_2MnTeO_6$, whereas the latter transition is from a minute $Mn_3O_4$ impurity not detectable by X-ray diffraction. $Mn_3O_4$ impurities are common when Mn perovskites are prepared in oxidizing conditions, which for $Ba_2MnTeO_6$ are needed in order to oxidize $Te^{4+}$ to $Te^{6+}$. The Néel temperature of $Ba_2MnTeO_6$ was determined from the first derivative of $\chi_{mol}T$ vs. T curve as $T_N$ = 20.3(2) K.[38]

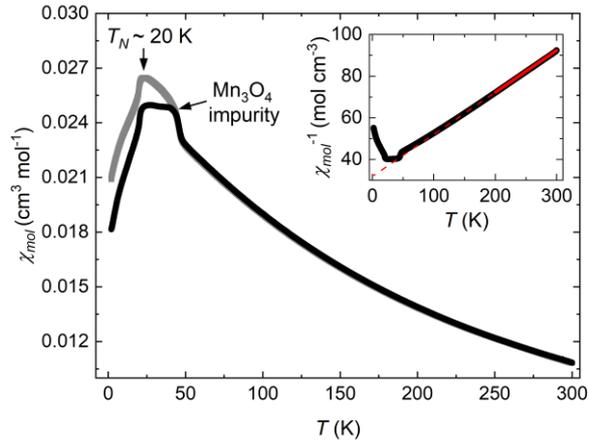

Figure 2. Magnetic susceptibility of $Ba_2MnTeO_6$. The zero-field cooled and the field cooled curves are shown in gray and black, respectively. An antiferromagnetic transition is observed at $T_N$ = 20 K, and a ferrimagnetic transition due to $Mn_3O_4$ impurity is observed at 45 K. Inset: Inverse magnetic susceptibility and a fit to the Curie-Weiss law yielding $\Theta_{CW}$ = -157(1) K and $\mu_{eff}$ = 6.30(1) $\mu_B$.

The inverse magnetic susceptibility was fitted to the Curie-Weiss law $\chi_{mol}$ = $C/(T-\Theta_{CW})$ in the temperature range 200-295 K. The Curie-Weiss constant $\Theta_{CW}$ gives an estimate of the overall strength of magnetic interactions in a material. For $Ba_2MnTeO_6$ we obtained $\Theta_{CW}$ = -157(1) K indicating strong antiferromagnetic interactions between the $3d^5$ $Mn^{2+}$ cations. The Curie constant was found to be C = 4.96(1) leading to an effective paramagnetic moment of $\mu_{eff}$ = 6.30(1) $\mu_B$. This is slightly larger than the expected spin-only value of $\mu_{SO}$ = 5.93 $\mu_B$ for S = 5/2 $Mn^{2+}$, but is similar to other reported $Mn^{2+}$ double perovskites.[23,39] The degree of magnetic frustration in $Ba_2MnTeO_6$ was estimated with the frustration factor f = $\Theta_{CW}/T_N$ ~ 8, which indicates that the compound is moderately but not strongly frustrated.[40]

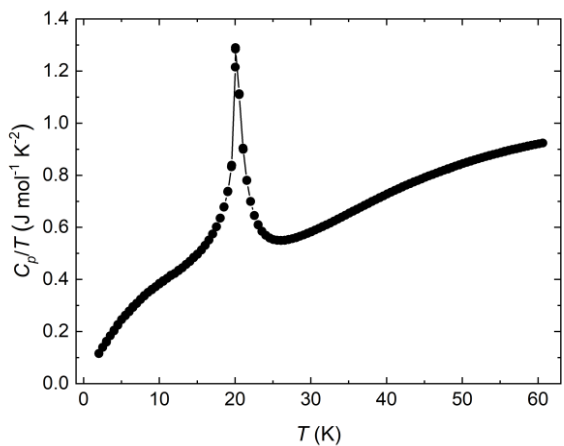

Figure 3. Specific heat of $Ba_2MnTeO_6$ between 2 K and 60 K. The λ anomaly observed at $T_N$ = 20 K is the only magnetic transition in $Ba_2MnTeO_6$ in this temperature range.

The specific heat of $Ba_2MnTeO_6$ was measured to clarify the magnetic transitions (Figure 3). A single $\lambda$ anomaly is observed at 20 K indicating a second order magnetic transition at $T_N$ = 20 K. This confirms that the second feature observed in magnetic susceptibility at 45 K is from a $Mn_3O_4$ impurity and not intrinsic to $Ba_2MnTeO_6$. While a small ferrimagnetic impurity can result in a large signal in the magnetic susceptibility, it will not be observed in the specific heat due to the very small change in the overall entropy of the sample.

Crystal structure

Crystal structures of perovskites can be predicted using the Goldschmidt tolerance factor calculated from the ionic radii of the different cations and the oxygen anion.[1] The tolerance factor for $Ba_2MnTeO_6$ is t = 1.00 corresponding to the cubic $Fm\bar{3}m$ structure also observed for $Ba_2MnWO_6$. $Te^{6+}$ double perovskites are generally known to follow the tolerance factor, although they are more likely to form hexagonal perovskite phases than their $W^{6+}$ analogues.[41] Wulff et al.[24] reported the crystal structure of $Ba_2MnTeO_6$ to be rhombohedrally distorted $R\bar{3}m$ and attributed the distortion to the different sizes of the $MnO_6$ and $TeO_6$ octahedra. However, their reported structure is cubic within one standard deviation for both the lattice parameters and atomic positions. To resolve this discrepancy, we performed neutron diffraction measurements at the GEM instrument at the ISIS Neutron and Muon Source.

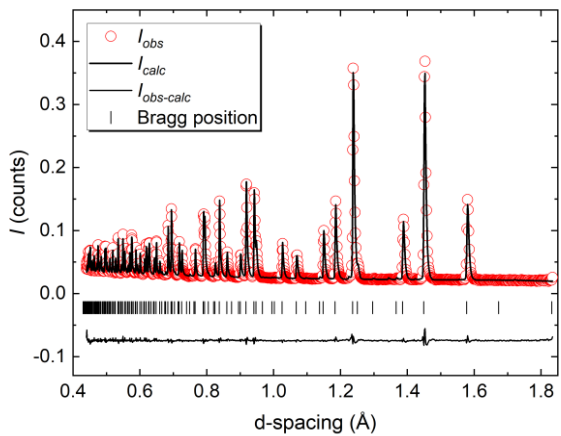

Figure 4. Rietveld refinement of the 100 K neutron diffraction data for $Ba_2MnTeO_6$ from the high-resolution bank 6 on GEM. The cubic $Fm\bar{3}m$ structure fits very well, and no features suggesting a lower symmetry structure are observed.

We found the crystal structure of $Ba_2MnTeO_6$ to be cubic based on both room-temperature X-ray diffraction and low-temperature (100 K) neutron powder diffraction measurements. All reflections could be indexed in the cubic $Fm\bar{3}m$ space group, and no peak splitting nor anisotropic line broadening expected for a small rhombohedral distortion were observed. The refinement of the 100 K neutron data on the high-resolution bank 6 of GEM is presented in Figure 4. The sample was found to be of high quality with a minor 1.0(1) wt% $BaMnO_3$ impurity, while the trace $Mn_3O_4$ impurity found in magnetometry could not be observed. The $Mn^{2+}$ and $Te^{6+}$ cations were found to be fully ordered on the two B-sites, and no antisite disorder was observed. This is expected as the B-site cations in $Ba_2MnTeO_6$ have a large charge difference, which drives the ordering.[1]

For comparison, the refined $Fm\bar{3}m$ structure was transformed to $R\bar{3}m$ using the WYCKSPLIT[42] and TRANSTRU tools provided by the Bilbao Crystallographic Server.[43–45] This space group allows for one additional degree of freedom in the lattice parameters, one in the position of barium and one in the position of oxygen. Rietveld refinement in $R\bar{3}m$ lead to only a negligible improvement with the additional positional and lattice parameters remaining within standard error of their cubic values. This confirms the structure is indeed cubic.

Table I. Refined structure of $Ba_2MnTeO_6$ at 2.1 K and 100 K. Space group Fm-3m with Ba on (1/4, 1/4, 1/4) site, Mn on (0, 0, 0) site, Te on (0, 0, 1/2) site and O on (x, 0, 0) site. The R-factors are reported for the high-resolution backscatter bank 6 (2θ = 153.90°).

|  | 2.1 K | 100 K |
|---|---|---|
| a (Å) | 8.2066(3) | 8.2106(4) |
| Ba 100 x $U_{ISO}$ (Å$^2$) | 0.01(4) | 0.13(4) |
| Mn 100 x $U_{ISO}$ (Å$^2$) | 0.00(4) | 0.03(4) |
| Te 100 x $U_{ISO}$ (Å$^2$) | 0.03(3) | 0.07(3) |
| O x | 0.2646(1) | 0.2647(1) |
| O 100 x $U_{ISO}$ (Å$^2$) | 0.22(2) | 0.32(2) |
| $BaMnO_3$ (%) | 1.0(1) | 1.0(1) |
| k | (0,0,1) | - |
| $m_{Mn}$ ($\mu_B$) | 4.34(3) | - |
| $R_p$ (%) | 6.61 | 7.44 |
| $R_{wp}$ (%) | 6.47 | 6.84 |
| $R_{exp}$ (%) | 1.89 | 3.50 |
| $\chi^2$ | 11.7 | 3.82 |

Magnetic structure

Magnetic Bragg peaks were observed in the neutron diffraction patterns below ~20 K. These were found to correspond to the propagation vector k = (0,0,1), which is equivalent to k = (1,0,0) = (0,1,0) in $Fm\bar{3}m$. This indicates Type I antiferromagnetic order in $Ba_2MnTeO_6$. The symmetry-allowed magnetic structures were evaluated using representation analysis with BASIREPS[27] and SARAh[46]. Two irreducible representations were found for the Mn (0,0,0) site in $Ba_2MnTeO_6$: $\Gamma_{mag} = \Gamma_7 + \Gamma_{10}$. The correct solution $\Gamma_{10}$ corresponds to ferromagnetic layers of $Mn^{2+}$ cations with in-plane moments stacked antiferromagnetically along c. The other solution $\Gamma_7$ has the Mn moment out of plane along c, but grossly fails to reproduce the main magnetic peak intensity. The direction of magnetic moment within the ab plane cannot be determined from powder data. We chose to set it arbitrarily along a. The combined nuclear and magnetic refinement at 2.1 K is shown in Figure 5a. The Type I magnetic structure fits the experimental data well. The higher $\chi^2$ for the 2.1 K refinement is due to an extended counting time compared to the 100 K dataset.[47]

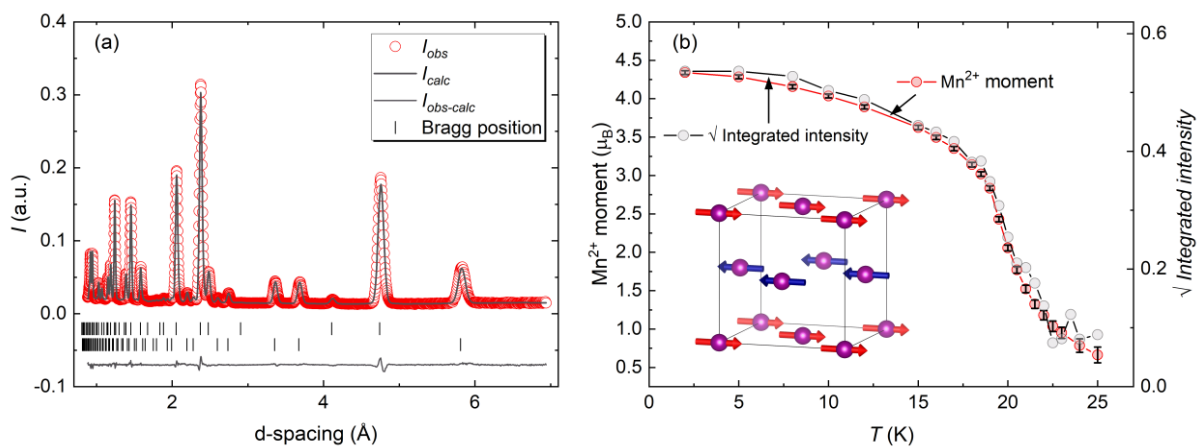

Figure 5. A) Rietveld refinement of the 2.1 K neutron diffraction data for $Ba_2MnTeO_6$. The upper bar symbols represent the nuclear phase, while the lower bar symbols mark the positions of the magnetic Bragg reflections. $Ba_2MnTeO_6$ has the Type I antiferromagnetic structure with a moment of 4.34(3) $\mu_B$ at 2.1 K. b) Thermal evolution of the refined magnetic moment (left) and the square root of the integrated intensity of the main magnetic peak (right). Inset: Magnetic structure of $Ba_2MnTeO_6$.

The refined magnetic moment of Mn at 2.1 K of 4.34(3) $\mu_B$ is as expected for S = 5/2 $Mn^{2+}$. We have plotted the refined moment as function of temperature in Figure 5b. At low temperatures the moment is reduced by spin waves, whereas the reduction in moment is governed by critical behavior near $T_N$. The refined moment does not decrease as fast as expected in the critical region. This is likely due to significant diffuse magnetic scattering, which is also observed above $T_N$. We have plotted the thermal evolution of the square root of the integrated intensity of the main magnetic ($\bar{1}10$) peak for comparison. Above 22 K the peak becomes indistinguishable from background, but broad diffuse magnetic scattering remains up to at least 40 K. The magnetic diffuse scattering suggests, that a short-range correlated magnetic state forms above $T_N$ similar to $Ba_2MnWO_6$.[23]

Magnetic interactions and excitations

Magnetic excitations in materials can be investigated using inelastic neutron scattering. For magnetically ordered materials, these excitations are spin waves. Magnetic exchange interactions can then be extracted from the spin wave spectra using linear spin wave theory.[48] Inelastic neutron scattering results for $Ba_2MnTeO_6$ are presented in Figure 6. The data collected at 7 K show clear spin wave excitations as expected in the magnetically ordered state. The spin wave spectra were simulated with SpinW.[48] A simple Heisenberg model was sufficient to describe the data:

$$H = -J_1 \sum_{(ij)} S_i \cdot S_j - J_2 \sum_{\langle ij \rangle} S_i \cdot S_j \qquad (1)$$

where $J_1$ is the nearest-neighbor interaction, $J_2$ is the next-nearest-neighbor interaction, $S_i$ is spin on site i and the sums are taken over their respective bonds. The spin wave simulation in Figure 6b reproduces the experimental spectra very well. This is also further supported by the vertical cuts shown in Figure 6c, in which the model closely follows the experimental data. We obtained $J_1$ = -0.34 and $J_2$ = 0.03 meV from the SpinW simulations consistent with the observed Type I antiferromagnetic structure. To conclude, $J_1$ is the main magnetic interaction in $Ba_2MnTeO_6$.

Unexpectedly some magnetic excitations persist far above $T_N$, as they are observed in the spectra collected at 44 K (Figure 6d) and 109 K (Figure 6e). The excitations get weaker with increased temperature and $|Q|$, which confirms their magnetic origin. The presence of magnetic excitations above $T_N$ indicates, that a short-range correlated state forms in $Ba_2MnTeO_6$. This is consistent with the significant diffuse magnetic scattering observed in the neutron diffraction experiments. This state survives up to at least 109 K, which is over 5 times $T_N$ = 20 K. In Figure 6f we present horizontal cuts through the data at different temperatures. The cuts show that the magnetic excitations above $T_N$ occur in the same positions in momentum transfer $|Q|$ as the spin waves in the ordered state. This suggests that the short-range correlated state up to $5T_N$ is closely related to the ordered magnetic state below $T_N$ = 20 K.

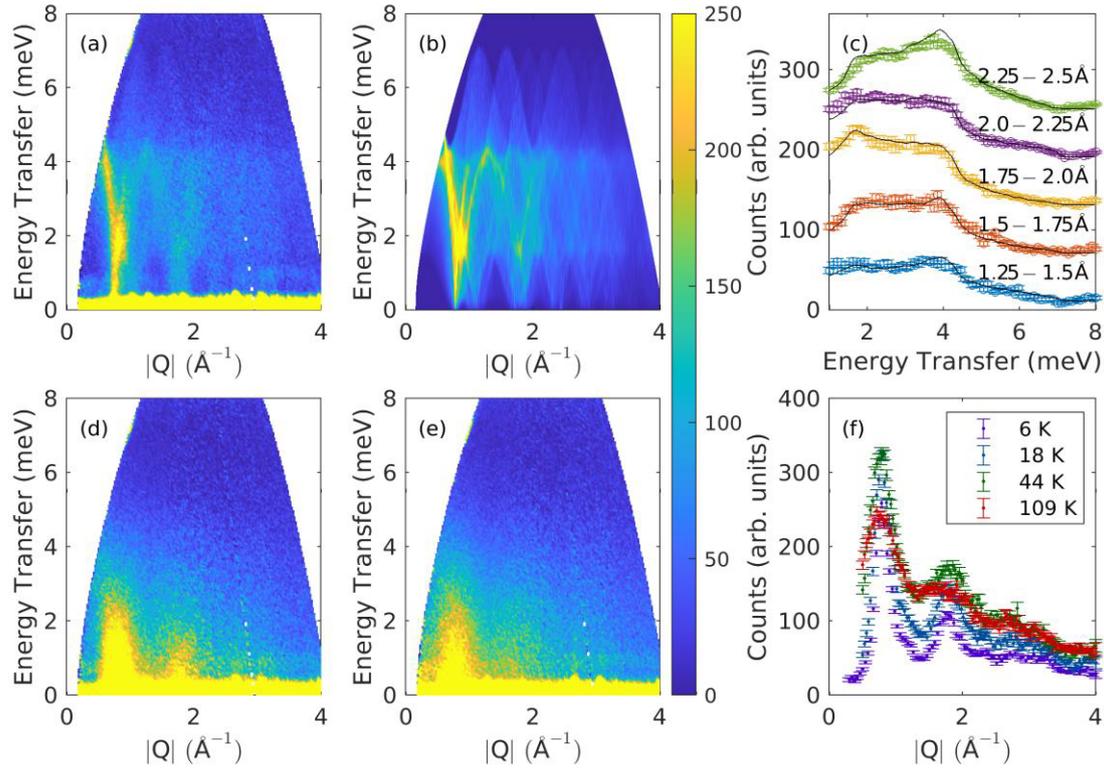

Figure 6. Inelastic neutron scattering from $Ba_2MnTeO_6$. a) 7 K data showing clear spin wave excitations. b) Spin wave simulation with $J_1$ = -0.34 meV and $J_2$ = 0.03 meV. c) Vertical cuts through the 7 K data summed over different |Q| intervals (squares) and the simulated spectra (black line) confirming that the simulation fits the data well. d) Data collected at 44 K still show magnetic excitations indicating a short-range correlated magnetic state. e) Magnetic excitations are observed even at 109 K. f) Horizontal cuts through the S(Q,E) data summed over E = 1–2 meV at different temperatures show that the magnetic excitations observed above $T_N$ = 20 K occur at the same positions in |Q| where the maxima in intensity are observed in the spin wave spectrum at 7 K.

Electronic structure

Density functional theory calculations in the DFT+U framework were used to investigate the electronic structures of $Ba_2MnTeO_6$ and $Ba_2MnWO_6$. $Ba_2MnTeO_6$ is an antiferromagnetic insulator as expected with an indirect band gap of $E_g$ = 1.75 eV. The partial density of states for $Ba_2MnTeO_6$ with U = 7 eV is shown in Figure 7a. The valence band below the Fermi level consists of hybridized Mn 3d and O 2p states. Similarly, in the conduction band the empty 3d Mn states hybridize with O 2p states. There is only weak hybridization of the empty Te 5s and 5p states with O 2p or Mn 3d states. The Te $4d^{10}$ states are located significantly below $E_F$ and are not shown. This suggests that tellurium does not participate in the extended superexchange interactions between the $Mn^{2+}$ cations.

Ba$_2$MnWO$_6$ is also an antiferromagnetic insulator with E$_g$ = 1.98 eV. The partial density of states, shown in Figure 7b, is different from Ba$_2$MnTeO$_6$ in important ways. The conduction band forms from unoccupied W 5d$^0$ states, which hybridize strongly with O 2p and, to a lesser degree, Mn 3d states. The valence band is formed of hybridized Mn 3d and O 2p states as in Ba$_2$MnTeO$_6$. The strong W 5d$^0$ – O 2p hybridization indicates that the empty tungsten 5d orbitals participate in the superexchange.

DISCUSSION

The typical magnetic structures for B-site ordered double perovskites are the Type I, Type II and Type III antiferromagnetic order derived from the ground states of the fcc Heisenberg model.[1,49] The Type I structure consists of alternating ferromagnetic layers stacked along [001]. This structure occurs when the antiferromagnetic J$_1$ interaction dominates, but it is frustrated as only 8 of the 12 nearest neighbor spins can couple antiferromagnetically. In the Type II structure, the alternating ferromagnetic layers are stacked along [111]. In this structure all the next-nearest-neighbor spins couple antiferromagnetically, and as such it requires a significant antiferromagnetic J$_2$ interaction. The type III structure is rare and a mixture of the first two.[1,50] Finally, Néel antiferromagnetic order is observed in some copper double perovskites. In these tetragonal compounds, interactions are strong in the ab-plane while out-of-plane interactions are negligible. In the Néel structure, all the nearest neighbor in-plane spins couple antiferromagnetically lifting the frustration found in the Type I structure. As such, it requires a dominant J$_1$ interaction.[49] All of these magnetic structures are visualized in Figure 8.

The cubic perovskites Ba$_2$MnTeO$_6$ and Ba$_2$MnWO$_6$ are isostructural with nearly identical lattice parameters and bond lengths. The only difference in the compounds is the presence of either 4d$^{10}$ Te$^{6+}$ or 5d$^0$ W$^{6+}$ on the B″-site, which links the extended superexchange pathways between the Mn$^{2+}$ cations. However, this minor difference is enough to result in completely different magnetic interactions and structures: Ba$_2$MnTeO$_6$ has the Type I structure with J$_1$ = -0.34 and J$_2$ = 0.03 meV, whereas Ba$_2$MnWO$_6$ has the Type II structure with J$_1$ = -0.080 and J$_2$ = -0.076. This major difference is explained by the differences in orbital hybridization observed in the density of states. The tellurium 5s and 5p orbitals do not significantly hybridize with O 2p and Mn 3d orbitals. The dominant J$_1$ interaction is a 90° Mn – O – Te – O – Mn superexchange, which can occur via the overlapping 2p orbitals of the two neighboring oxygens without participation of tellurium. The J$_2$ interaction occurs along a 180° Mn – O – Te – O – Mn superexchange pathway, which must involve the bridging tellurium as the oxygens are too far away for their orbitals to overlap. Thus, the limited hybridization of Te with O 2p explains the weak $J_2$ interaction in $Ba_2MnTeO_6$. In $Ba_2MnWO_6$, the empty $5d^0$ W orbitals hybridize strongly with O 2p. This hybridization enables the significant $J_2$ superexchange via the bridging W cations. Thus, the differences in the magnetic interactions of $Ba_2MnTeO_6$ and $Ba_2MnWO_6$ can be explained with a $d^{10}/d^0$ orbital hybridization effect[5] as has been reported for some copper double perovskites.[11–14]

It is perhaps surprising that $Ba_2MnTeO_6$ has stronger magnetic interactions and a higher $T_N$ than $Ba_2MnWO_6$, since in the former the tellurium hybridizes with O/Mn only weakly whereas in the latter the W hybridizes and participates in the superexchange much more strongly. This is not a structural effect as $Ba_2MnWO_6$ has a slightly smaller lattice parameter, where one would expect slightly stronger magnetic interactions as a result of increased orbital overlap. In $Sr_2CuWO_6$, it was found that the W $5d^0$ – O 2p hybridization also suppresses the $J_1$ interaction in addition to promoting $J_2$.[13] While this is not necessarily directly translatable to $Ba_2MnWO_6$ due to the different 3d orbitals involved on the magnetic cation, this could very well be the origin of the weaker interactions in $Ba_2MnWO_6$.

For comparison, in Table 2 we have collected crystallographic and magnetic data for $A_2MB''O_6$ double perovskites, where M is a divalent 3d transition metal cation and B'' is $Te^{6+}$ or $W^{6+}$. We have only included compounds where both the $Te^{6+}$ and $W^{6+}$ analogues form an ordered double perovskite. The monoclinic $P2_1/n$ $Sr_2MnTeO_6$ and $Sr_2MnWO_6$ behave similarly to the barium analogues such that the former has the Type I structure while the latter has Type II.[39,51,52] All the compounds with a $d^{10}$ $Te^{6+}$ linking B'' cation have either Type I or Néel antiferromagnetic structures, revealing dominant antiferromagnetic $J_1$ interactions. In contrast, all the $d^0$ $W^{6+}$ analogues have Type II antiferromagnetic order indicating significant antiferromagnetic $J_2$ interactions. While we have limited our discussion here to $Te^{6+}$ and $W^{6+}$ double perovskites due to their near-identical ionic radii, the same Type II antiferromagnetic structure observed in $5d^0$ $W^{6+}$ compounds is also found in the corresponding $4d^0$ $Mo^{6+}$ materials including $Ba_2MnMoO_6$.[6,52–56] This further supports that the differences in magnetic structures and interactions are driven by a $d^{10}/d^0$ effect.

We can make two main conclusions based on the observed magnetic structures. 1) $d^{10}$ cations on the B''-site promote strong antiferromagnetic $J_1$ interactions leading to Type I or Néel magnetic structures. The Type I antiferromagnetic structure is only possible when $J_1$ is antiferromagnetic and $J_2$ is ferromagnetic. Since the Curie-Weiss constant, $\Theta_{CW}$, is negative for all listed $Te^{6+}$ compounds, the antiferromagnetic $J_1$ is much stronger than the ferromagnetic $J_2$ in these compounds. 2) $d^0$ cations on the B''-site promote antiferromagnetic $J_2$ interactions leading to Type II magnetic structures such that $J_2/J_1 > 0.5$, while $J_1$ also remains antiferromagnetic. For some Type II compounds where the interactions are known, the antiferromagnetic $J_2$ is dominant; but for $Ba_2MnWO_6$ it is of the same order as $J_1$. These rules open up the possibility of tuning magnetic interactions in double perovskites by making substitutions on the B″-site. This effect has already been demonstrated in the $Cu^{2+}$ double perovskite series $Sr_2CuTe_{1-x}W_xO_6$, where the ground state can be tuned from Néel order to a spin-liquid-like state to Type II order as function of x.[14,15]

It is not presently known how widely the $d^{10}/d^0$ rules established here are applicable to other B-site ordered transition metal double perovskites with $d^{10}$ and $d^0$ B″ cations. It should be noted that other factors such as the degree of cation ordering and octahedral tilting also affect the magnetism in double perovskites. $Sr_2FeSbO_6$ with partial cation ordering of $Fe^{3+}$ and the $d^0$ $Sb^{5+}$, for instance, does have the expected Type I structure, but the B-sites are disordered in the $Nb^{5+}$ and $Ta^{5+}$ $d^0$ analogues leading to spin glass states.[57,58] Similarly, $Sr_2CrSbO_6$ is a Type I antiferromagnet while the more distorted $Ca_2CrSbO_6$ with larger octahedral tilting becomes ferromagnetic.[59] $SrLaNiSbO_6$ and $BaLaCuSbO_6$ have the expected Type I and Néel magnetic structures, respectively.[60,61] Differences between $d^{10}$ $Zn^{2+}$ and $Cd^{2+}$ and $d^0$ $Ca^{2+}$ $Ba_2B'OsO_6$ $Os^{6+}$ ($5d^2$) compounds have also been reported, but the magnetic structures are not known.[18] Curiously, the $Os^{5+}$ ($5d^3$) perovskites $Sr_2B'OsO_6$ with $d^{10}$ $Sc^{3+}$ and $In^{3+}$ and $d^0$ $Y^{3+}$ B′ cations all share the Type I antiferromagnetic structure, although unexpected differences in the exchange interactions strengths have also been attributed to a $d^{10}/d^0$ orbital hybridization effect.[19–21]

Table II. Ordered II-VI double perovskites with $d^{10}$ ($Te^{6+}$) or $d^0$ ($W^{6+}$) B″ cations and their magnetic properties.

| Compound | Space group (low T) | $T_N$ (K) | Magnetic structure | $\Theta_{CW}$ (K) | $J_1$ (meV) | $J_2$ (meV) | $J_2/J_1$ | Ref |
|---|---|---|---|---|---|---|---|---|
| $Ba_2MnTeO_6$ | $Fm\bar{3}m$ | 20 | Type I | -157 | -0.34 | 0.03 | -0.09 | * |
| $Sr_2MnTeO_6$ | $P2_1/n$ | 20 | Type I | -136 | - | - | - | 39 |
| $Sr_2CoTeO_6$ | $P2_1/n$ | 18 | Type I | -140 | - | - | - | 62 |
| $Sr_2NiTeO_6$ | $P2_1/n$ | 35 | Type I | -225 | - | - | - | 63,64 |
| $Ba_2CuTeO_6$ | $I4/m$ | - | Néel | -400 | -20.22 | 0.23 | -0.01 | 12,17 |
| $Sr_2CuTeO_6$ | $I4/m$ | 29 | Néel | -80 | -7.18 | -0.21 | 0.03 | 9,10,14 |
| $Ba_2MnWO_6$ | $Fm\bar{3}m$ | 8 | Type II | -63 | -0.080 | -0.076 | 0.95 | 23 |
| $Sr_2MnWO_6$ | $P2_1/n$ | 14 | Type II | -71 | - | - | - | 51,52 |
| $Sr_2CoWO_6$ | $P2_1/n$ | 24 | Type II | -62 | - | - | - | 65 |
| $Sr_2NiWO_6$ | $I4/m$ | 54 | Type II | -175 | -0.02 | -1.81 | 90 | 64,66,67 |
| $Ba_2CuWO_6$ | $I4/m$ | 28 | Type II | -249 | -1.17 | -11.94 | 10.18 | 12,17,68 |
| $Sr_2CuWO_6$ | $I4/m$ | 24 | Type II | -165 | -1.2 | -9.5 | 7.92 | 6–8,14 |

CONCLUSIONS

We have characterized the magnetic properties of the B-site ordered double perovskite $Ba_2MnTeO_6$. It has a cubic $Fm\bar{3}m$ structure with a = 8.2066(3) Å at 2 K. The compound orders magnetically at $T_N$ = 20 K in the Type I antiferromagnetic structure, and has a Curie-Weiss temperature of -157(1) K indicating significant antiferromagnetic interactions. The exchange constants were extracted from inelastic neutron scattering data. Magnetism in $Ba_2MnTeO_6$ is well described by a fcc Heisenberg model with two interactions: nearest-neighbor $J_1$ = -0.34 meV and next-nearest-neighbor $J_2$ = 0.03 meV. A short-range correlated magnetic state was observed at high temperatures up to T ~5$T_N$.

$Ba_2MnTeO_6$ and its tungsten analogue $Ba_2MnWO_6$ are excellent materials for testing the effects of filled shell $d^{10}$ ($Te^{6+}$) and empty shell $d^0$ ($W^{6+}$) bridging B″-site cations on magnetic interactions in perovskites. The compounds are isostructural with very similar lattice parameters due to the similar ionic radii of $Te^{6+}$ and $W^{6+}$. The magnetic structures and interactions in these compounds are very different: $Ba_2MnTeO_6$ has the Type I structure with a dominant $J_1$ interaction, while $Ba_2MnWO_6$ has the Type II structure with a significant antiferromagnetic $J_2$ interaction. This arises due to differences in the orbital hybridization with oxygen. The empty W $d^0$ states hybridize strongly with oxygen 2p states, and consequently tungsten participates in the next-nearest-neighbor extended superexchange pathway enabling a significant $J_2$ interaction. In contrast, the filled $d^{10}$ Te states are far below the Fermi level, while the empty s and p orbitals hybridize only weakly with O 2p. This leads to a very weak $J_2$ interaction, while $J_1$ dominates. This $d^{10}/d^0$ effect, where $d^{10}$ cations in the bridging B″ sites promote $J_1$ interactions and $d^0$ B″ cations promote $J_2$ interactions, is also observed in other B-site ordered double perovskites. Magnetic interactions in double perovskites could be tuned by substituting $d^{10}/d^0$ cations on the B″-site.

Acknowledgements

This work was funded by the Leverhulme Trust Research Project Grant RPG-2017-109. Dr Ivan da Silva is thanked for assistance with the neutron diffraction measurements. Dr Alex Gibbs is thanked for helpful discussions. The authors are grateful to the Science and Technology Facilities Council for the beamtime allocated at ISIS. The authors are thankful for access to the MPMS3 and PPMS instruments at the Materials Characterisation Laboratory at ISIS.

Otto H. J. Mustonen,[1] Charlotte E. Pughe,[1] Helen C. Walker,[2] Heather M. Mutch,[1] Gavin B. G. Stenning,[2] Fiona C. Coomer,[3] Edmund J. Cussen[1]*

[1]Department of Material Science and Engineering, University of Sheffield, Mappin Street, Sheffield S1 3JD, United Kingdom
[2]ISIS Pulsed Neutron and Muon Source, STFC Rutherford Appleton Laboratory, Harwell Campus, Didcot OX11 0QX, United Kingdom
[3]Johnson Matthey Battery Materials, Blount's Court, Sonning Common, Reading RG4 9NH United Kingdom

Table S1. Conventional R-factors for the Rietveld refinement of the neutron diffraction data at 2 K and 100 K.

|  | $R_p$ (%) | $R_{wp}$ (%) | $R_{exp}$ (%) | $R_p$ (%) | $R_{wp}$ (%) | $R_{exp}$ (%) |
|---|---|---|---|---|---|---|
| Bank 2 (2θ = 17.95°) | 9.37 | 8.63 | 5.03 | 13.4 | 10.3 | 10.23 |
| Bank 3 (2θ = 34.96°) | 4.92 | 5.05 | 2.26 | 6.37 | 6.04 | 4.08 |
| Bank 4 (2θ = 63.99°) | 4.53 | 5.04 | 1.27 | 4.27 | 4.86 | 2.27 |
| Bank 5 (2θ = 91.50°) | 6.18 | 7.22 | 1.40 | 6.02 | 7.08 | 2.50 |
| Bank 6 (2θ = 153.90°) | 6.61 | 6.47 | 1.89 | 7.44 | 6.84 | 3.50 |
| $\chi^2$ (global) | 12.5 | | | 4.13 | | |

Table S2. Bond valence calculations for the refined structure of Ba$_2$MnTeO$_6$ at 2 K. The calculations agree well with the expected oxidation states Ba$^{2+}$, Mn$^{2+}$ and Te$^{6+}$.

|  | Ba | Mn | Te |
|---|---|---|---|
| M-O (Å) | 2.9039(1) | 2.1711(8) | 1.9322(8) |
| BVS | 2.253 | 2.142 | 5.759 |

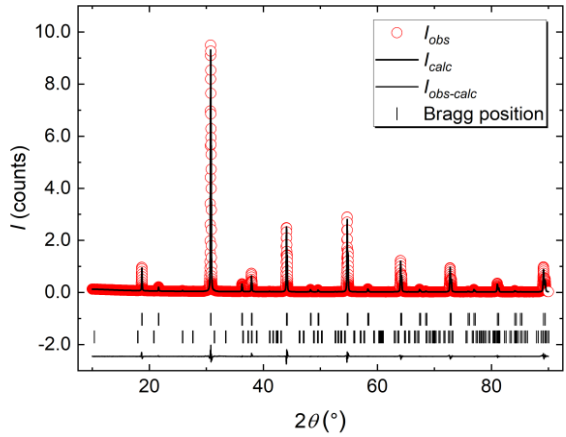

Figure S1. Rietveld refinement of laboratory x-ray data for Ba$_2$MnTeO$_6$ (Cu Kα source).

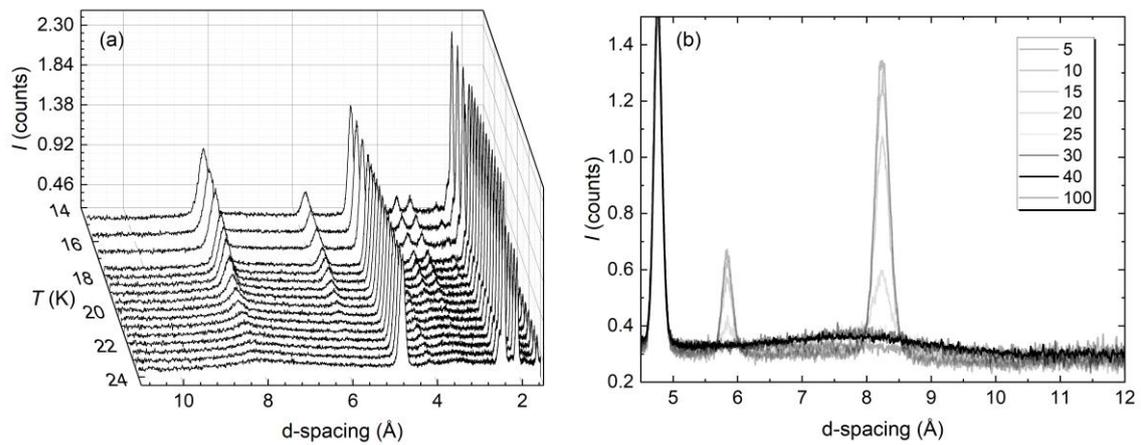

Figure S2. Low-temperature neutron diffraction of Ba$_2$MnTeO$_6$. (a) Thermal evolution of the magnetic Bragg reflections. (b) Neutron diffraction around the main magnetic Bragg peaks at different temperatures. Above T$_N$ ~ 20 K, significant diffuse magnetic scattering is observed as a wide bump around the main magnetic peaks.

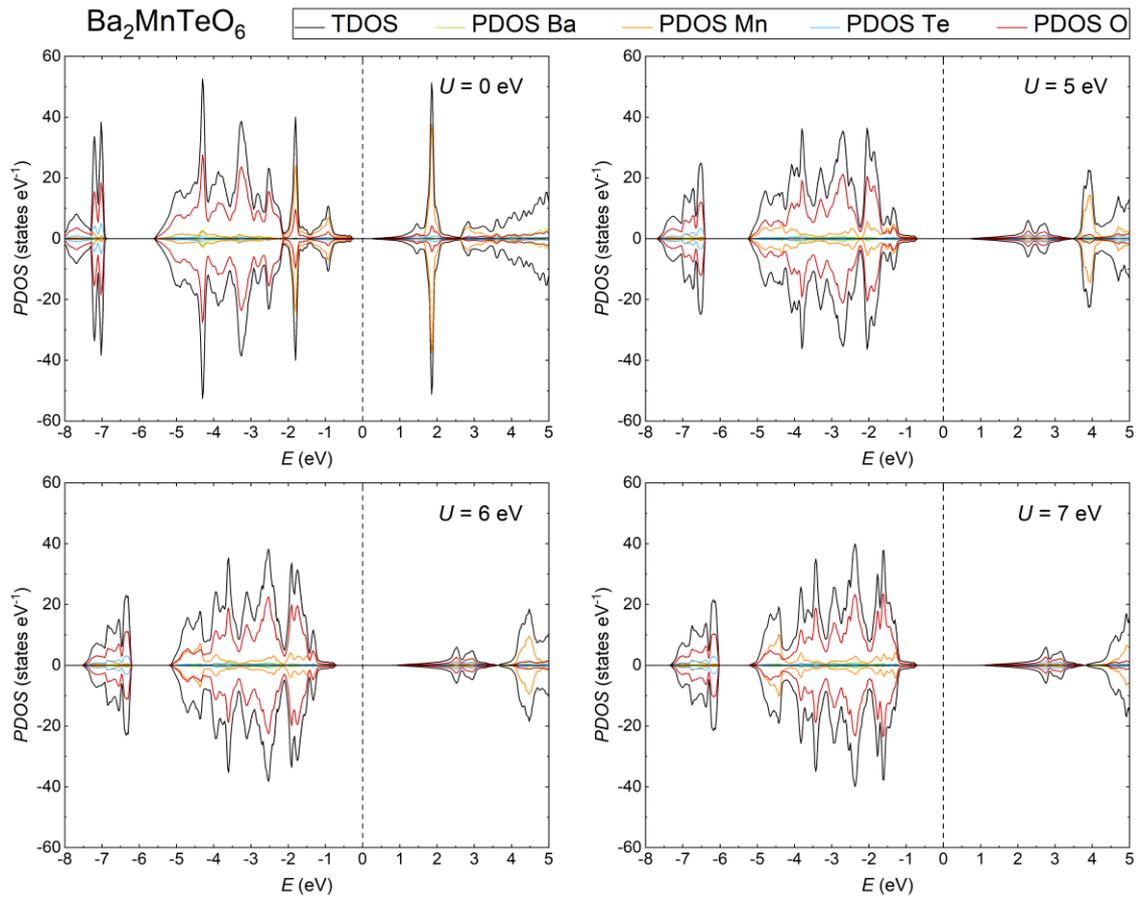

Figure S3. Calculated total and partial density of states for $Ba_2MnTeO_6$ with U = 0 (GGA-PBE), 5, 6 and 7 eV on the Mn 3d states. The band gap is significantly underestimated without U as is typical for 3d transition metal compounds. The calculations with U are all in good agreement and the main effect is the widened band gap. The Te states hybridize only lightly with O and Mn states.

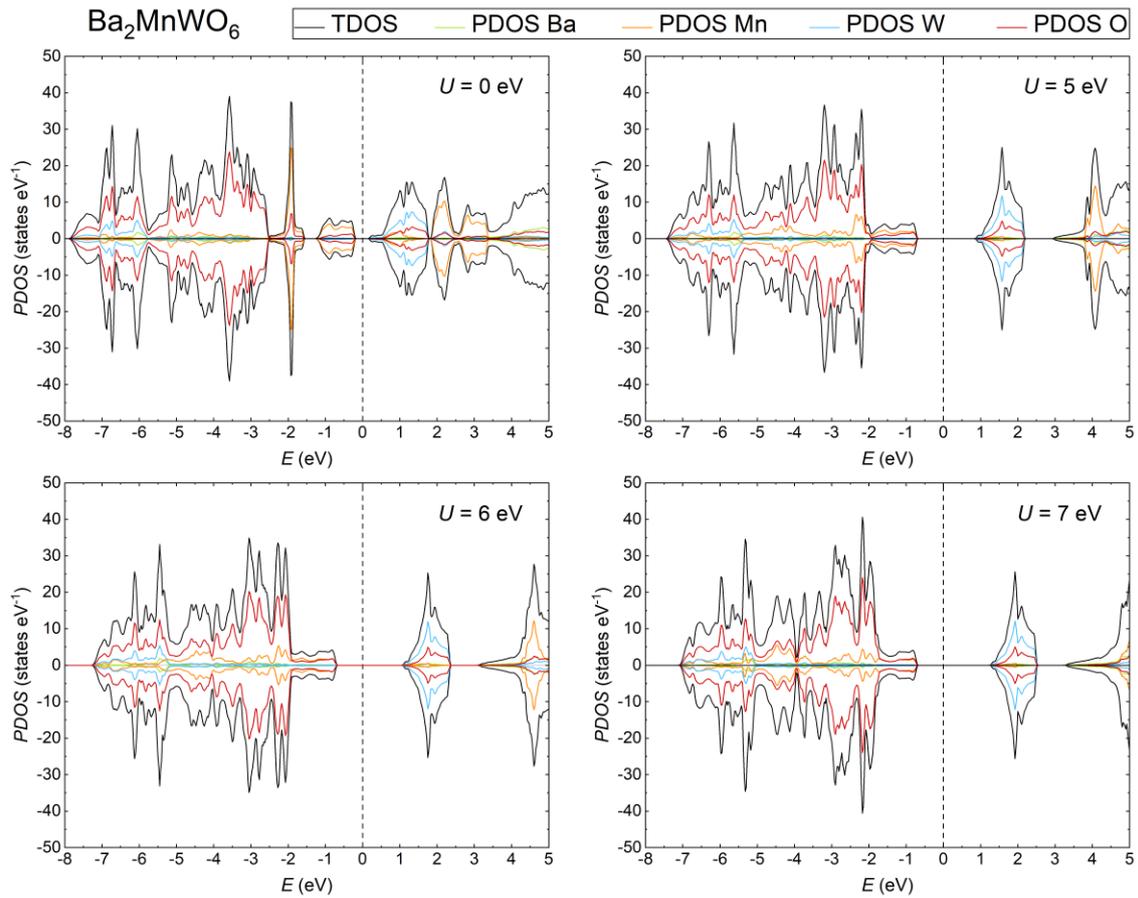

Figure S4. Calculated total and partial density of states for Ba$_2$MnWO$_6$ with U = 0 (GGA-PBE), 5, 6 and 7 eV on the Mn 3d states. The band gap is significantly underestimated without U as is typical for 3d transition metal compounds. The calculations with U are all in good agreement and the main effect is the widened band gap. The W states hybridize with O and Mn states especially in the conduction band.